\documentclass{aastex63}
\usepackage{amsmath}
\usepackage{wasysym}
\usepackage{wrapfig}
\usepackage{graphicx}
\usepackage{hyperref}
\usepackage{natbib}
\usepackage{float}
\usepackage{ulem}
\usepackage{xcolor}
\usepackage[caption=false]{subfig}

\begin{document}
\title{What Does It Mean to be a ``Depleted" Comet? High Spectral Resolution Observations of the Prototypical Depleted Comet 21P/Giacobini-Zinner from McDonald Observatory}
\shorttitle{High Spectral Resolution Observations of Giacobini-Zinner}
\author[0000-0003-4828-7787]{Anita L. Cochran}
\affil{McDonald Observatory, The University of Texas at Austin, Austin, TX, USA}
\author{Tyler Nelson}
\affil{Astronomy Department, The University of Texas at Austin, Austin, TX, USA}
\author[0000-0002-0622-2400]{Adam J. McKay}
\affil{NASA Goddard Space Flight Center, Greenbelt, MD, USA}
\affil{Dept. of Physics, American University, Washington DC USA }

\shortauthors{Cochran, Nelson, McKay}
\correspondingauthor{Anita Cochran}
\email{anita@astro.as.utexas.edu}

\begin{abstract}
We present high spectral resolution optical observations of comet 21P/Giacobini-Zinner from six nights in 2018. The observations were obtained with the Tull coud\'{e} spectrograph on the 2.7m Harlan J. Smith Telescope of McDonald Observatory.  This comet's spectrum shows strong depletions in C$_2$, C$_3$, CH, and NH$_2$ relative to CN. We explore what it means for a comet to be depleted and show that all of the species are present in the spectrum at similar relative line intensities within a a molecular band compared with a typical comet.  The depletions represent a much lower abundance of the species studied.

\end{abstract}

\section{Introduction}
Comets are some of the least altered bodies left over from the formation of the Solar System.  Made of dust and ice, these bodies spend much of their lives far from the Sun until some gravitational perturbation causes them to enter the inner Solar System.  Once there, the ices sublime. The gas flows away from the small, and gravitationally weak, nucleus, entraining dust in its flow.

Cometary spectra have been obtained for over 150 years and it  has long been noted that the spectra of most comets are remarkably similar. Photometric and spectroscopic surveys have been undertaken to assess if {\it all} comets are spectrally similar or whether there are comets with different compositions \citep{ahetal95,fi09,lasm11,coetal2012,coreview2015}. Based on production rate ratios from optical observations, these surveys found that approximately 75\% of the observed comets had very similar compositions, termed ``typical".  The other 25\% were depleted in C$_2$ and are designated ``depleted" comets.  All of the surveys use Q(C$_2$/OH) or Q(C$_2$/CN) to define typical versus depleted, where Q is the production rate in molecules sec$^{-1}$.  \citet{coetal2012} and \citet{ahetal95} also observed C$_3$. Using a very strict definition of C$_3$ depletion, Cochran et al. found that about 9\% of all observed comets were depleted in both C$_2$ and C$_3$.  With a slightly less strict definition, A'Hearn at al. found about 24\% of all comets were depleted in both C$_2$ and C$_3$. The C$_3$ band was not contained in the spectra of \citet{fi09} and \citet{lasm11}.  

\begin{figure}[h!]
	\centering
	\includegraphics[width=\linewidth,keepaspectratio]{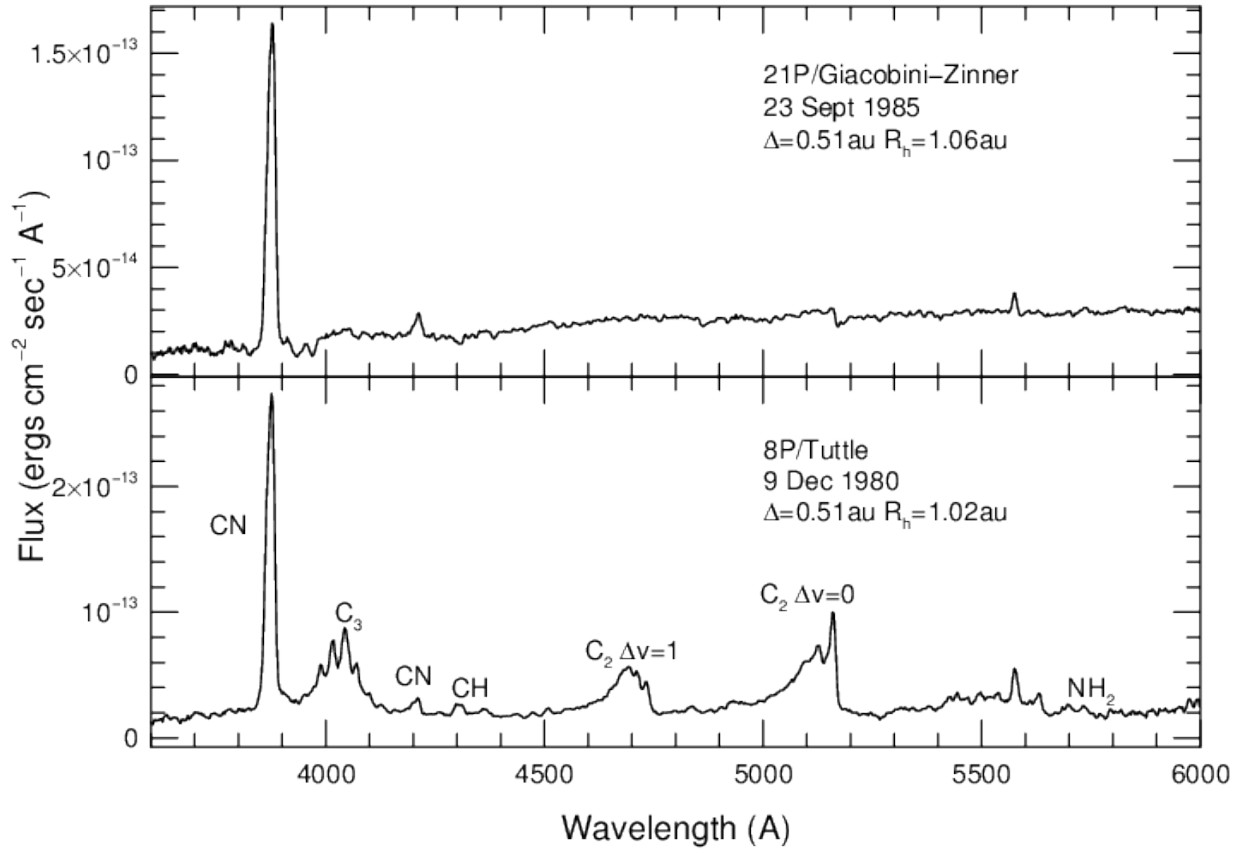}
	\caption{Spectra in the optical of GZ (top) and Tuttle (bottom)  are shown (Spectra from \citet{coetal2012}, figure 13).  Emission due to CN is prominent in both spectra.  However, while Tuttle shows strong molecular emissions due to C$_3$ and C$_2$, those features appear  extremely weak in the GZ spectrum. \label{lowres}}
\end{figure}

Comet 21P/Giacobini-Zinner (hereafter GZ) is the prototype of the depleted comets because it was the first one discovered \citep{scmibi87,coba87gz,bewasclu90,laetal03,fi09}; it is part of the group that are depleted  in both C$_2$ and C$_3$ in the definitions of both Cochran et al. and A'Hearn et al. Figure~\ref{lowres} shows a low resolution optical spectrum of GZ, along with a spectrum from the same instrument of typical comet 8P/Tuttle.  The two comets were at essentially the same heliocentric and geocentric distances at the times of the observations.  Inspection of these spectra show a prominent emission at $\sim3880$\AA\ due to CN in both spectra.  Tuttle also shows strong emissions of C$_2$ and C$_3$ that are very weak in the GZ spectrum.

Comparison of the spectra in Figure~\ref{lowres} raises the question of what it means for a comet to be depleted in a species.   Does this mean that there is some of that species, but the distribution of the relative line strengths in the spectrum is very different from a typical comet?  Or does it mean that all the same lines that we observe in a typical comet's spectrum are present with the same relative strengths as a typical comet, but that they are all just much weaker than we would expect for a typical comet? In order to answer this question, we undertook high spectral resolution observations of GZ in 2018. 

\section{Observations}

 High spectral resolution observations are more sensitive to very weak emission features, such a those found in comets, because, for a given equivalent width, a line subtends fewer pixels and  is more prominent above the continuum. Therefore, it is easier to detect a weak feature with higher spectral resolution.  We took advantage of the relatively close passage of GZ to the Earth in 2018 to obtain observations of GZ with the Tull coud\'e spectrograph \citep{TuMQSn95} on the 2.7m Harlan J. Smith Telescope of McDonald Observatory.  Table~\ref{log} shows details of our observations. 

\begin{table}[h!]
	\centering
	\caption{Log of Observations \label{log}}
	\begin{tabular}{l|r@{ }l@{ }l@{ } | c | c | c | c | c | l}
		\hline
	&	\multicolumn{3}{c|}{Date} & Heliocentric & Heliocentric & Geocentric & Geocentric & Total & \multicolumn{1}{c}{Sky} \\
	&	\multicolumn{3}{c|}{(UT)} & Distance & Velocity & Distance & Velocity & Exposure (s) & \multicolumn{1}{c}{Conditions} \\
	& \multicolumn{3}{c|}{} & (AU) & (km/s) & (AU) & (km/s) & (s) & \\
		\hline
	GZ&	2018 & Jul & 29 &1.18 & -12.17 & 0.62 & -13.60 & 4800 & Thin clouds \\
	&	2018 & Jul & 30 &1.18 & -11.99 & 0.61 & -13.42 & 6000 & Mostly clear\\
	&	2018 & Aug &  1 &1.16 & -11.63 & 0.60 & -13.33 &22800 & Very cloudy\\
	&	2018 & Aug & 18 &1.06 &  -7.66 & 0.47 & -10.59 &14400 & Very cloudy \\
	&	2018 & Aug & 20 &1.06 &  -7.08 & 0.46 & -10.04 &18000 & Very cloudy\\
	&	2018 & Sep & 15 &1.02 &   1.67 & 0.40 &   2.64 & 5700 & Clear\\
		\hline
Tuttle & 2007 & Dec & 27 & 1.14 & -10.78 & 0.27 & -12.57 & 6600 & Clear \\
		\hline
	\end{tabular}
\end{table}
		
The observations were obtained with a resolving power R=$\lambda/(\Delta\lambda)=60,000$. The instrument is a cross-dispersed echelle spectrograph.  The obtained spectra contained 65 spectral orders per readout, covering the range from 3700\AA-5700\AA\ completely and out to 1.02$\mu$m with increasing interorder gaps. The slit subtended 1.2\,arcsec $\times$ 8.2\,arcsec on the sky.  The slit is fixed in orientation and location in the dome, while the telescope rotates around the polar axis to track the object. Thus, the image rotates with respect to the slit throughout the night, covering 360$^\circ$ in 24 hours.  In the course of each night's observations, the slit would image slightly different position angles of the coma, but remained centered on the optocenter. For example, on 2018 Sep 15, the image rotated by 37.5$^\circ$ during the observations.  We cannot be sure how much of the cometary rotation period was covered during this rotation because the rotation period is unknown. The coma morphology evolved very slowly over the apparition. There is slight evidence that the period might be a whole number of days (2.0, 3.0), but it could be very long (Schleicher, personal communication).

We needed a relatively long exposure to achieve the high signal/noise required for studying details of the molecular bands. We obtained the cometary spectra using 20--30-minute exposures and we combined multiple exposures each night into a final spectrum. The use of multiple exposures allows us to filter out defects, such as radiation events, that hit the detector.   We combined the spectra in their raw image form, using a median combine after scaling by the average counts in a particular section of the image (the scaling accounts somewhat for any clouds). This combination process removes the radiation events in the most accurate way possible, as they show up in different locations in different spectra. After the combination, we renormalized to the equivalent total exposure time. 

The resultant spectra consist of three components: 1) molecular emissions from the cometary gas; 2) continuum and absorption features from the reflection of the solar spectrum off dust in the coma; and 
3) telluric lines (emissions and absorptions) superposed from the light passing through the Earth's atmosphere.  The third component only affects some of the orders,  though none of the orders discussed in this paper. The first two are seen in all orders.  With the small slit, very little sky spectrum is imaged onto the detector and any that is detected is removed along with the solar absorptions. Many zero second exposures (bias frames) were obtained each night to measure the bias of the chip.  Incandescent light spectra were obtained through the slit to remove the pixel-to-pixel variations (flat fields).  

Once bias and flat fields were removed from the spectral image, we traced the path of the spectrum on the chip and extracted individual orders by collapsing all of the pixels perpendicular to the trace.  For GZ, this was done on the combined nightly image.
Observations of a ThAr lamp were obtained each night to determine the wavelength of each point in the extracted spectrum, resulting in a known wavelength accuracy of about 24 m\AA.  

\section{Comparison with a ``typical" comet}
In order to determine what it means to be a depleted species, we needed to compare our spectra of GZ with another comet observed with the Tull spectrograph under comparable orbital circumstances.   We used observations of comet 8P/Tuttle obtained with the Tull spectrograph on 2007 December 27 UT, when the comet was at a heliocentric distance of 1.14\,AU and a geocentric distance of 0.27\,AU.  While not as perfect a match for heliocentric and geocentric distances as our comparison in Figure~\ref{lowres}, the Tuttle observations were adequate for our purposes.  For our comparisons, we used the GZ data from 2018 September 15 since that was the one night that was completely clear.  For all molecules except CH (see discussion below), the GZ spectra from all nights looked essentially the same.  We inspected all of the images of GZ obtained on 2018 Sep 15, when it was clear, to determine if there were any variations in the spectra during the course of the night. None were detected.

\begin{figure}[h!]
	\centering
	\includegraphics[scale=0.65]{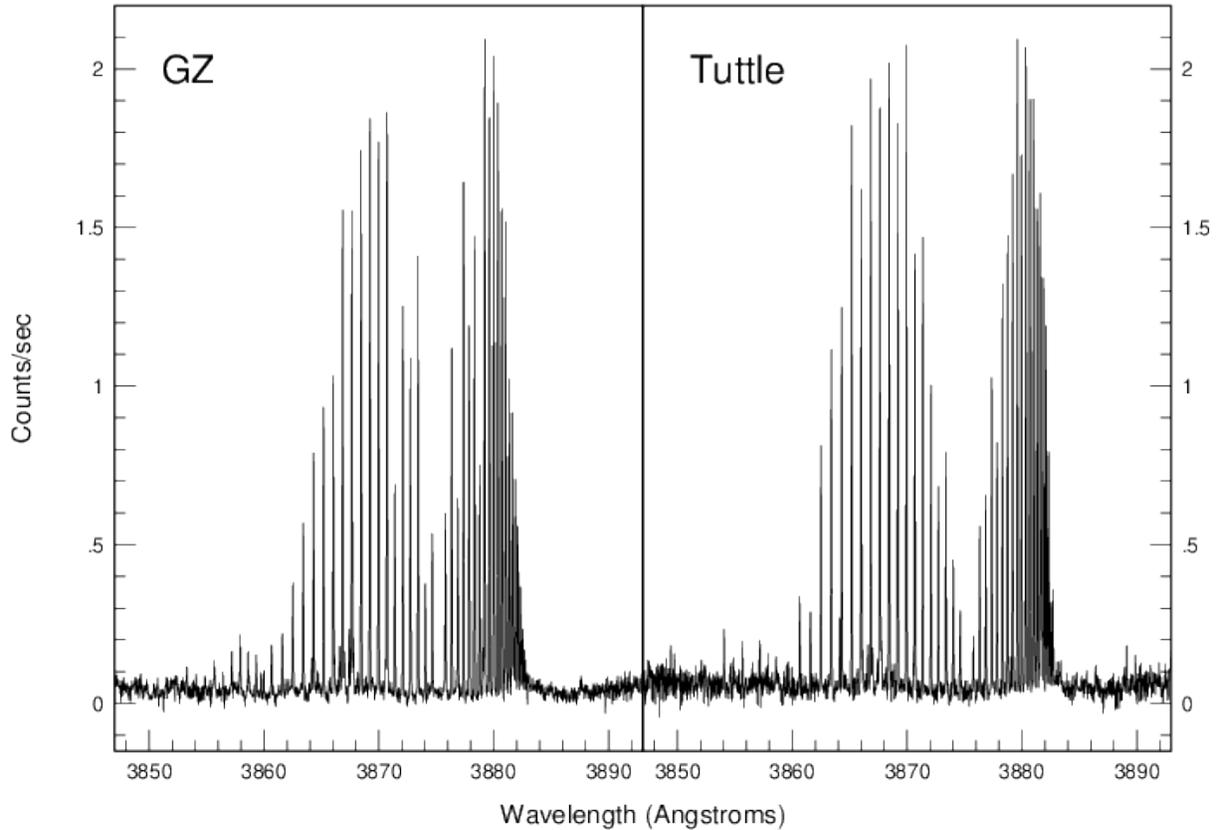}
	\caption{The CN spectra of GZ (left panel) from 2018 September 15 and Tuttle (right panel) from 2007 December 27 are shown.  The count rate is  similar for both comets.  The differences in the relative strengths of the lines can be attributed to differences in the Swings effect \citep{swingseffect} and rotational temperature of the gas.  \label{CN}}
\end{figure}

Figure~\ref{CN} shows the spectra of GZ from 2018 September 15 and Tuttle in the region of the CN B\,$^2\Sigma^+$ -- X\,$^2\Sigma^+$ band. For GZ, CN was by far the strongest band that we detected. The observations of GZ and Tuttle had different exposure times, but the  count rate (counts/second) at the strongest peaks were comparable in the two spectra. Examination of the two spectra show that they have comparable signal/noise.  The differences in the strengths of the lines, including the weaker R-branch relative to the P-branch in GZ, can be attributed to differences in the Swings effect \citep{swingseffect} and to the different heliocentric distance, thus gas temperature.

For a rigorous comparison based on the taxonomy of \citet{ahetal95}, the determination of whether a comet is typical or depleted  should rest on a comparison of the content of C$_2$ to that of H$_2$O, generally measured in the optical by observations of OH.  However, the OH (0,0) band is at 3080\AA, a wavelength not in the bandpass of the Tull spectrograph.   Therefore, we used count rates of species relative to the count rates in the CN band for determination of depletion. However, as both Tuttle and GZ are in the A'Hearn et al. taxonomy,  we know the CN/OH ratios in both comets.  CN/OH=0.18\% for GZ, while Tuttle has CN/OH=0.29\%, about 60\% higher.  While these are not identical,  and it could be accounted for in order to compare to OH, this difference is smaller than the depletions we observed in GZ.

\begin{figure}[h!]
	\centering
	\includegraphics[width=\linewidth,keepaspectratio]{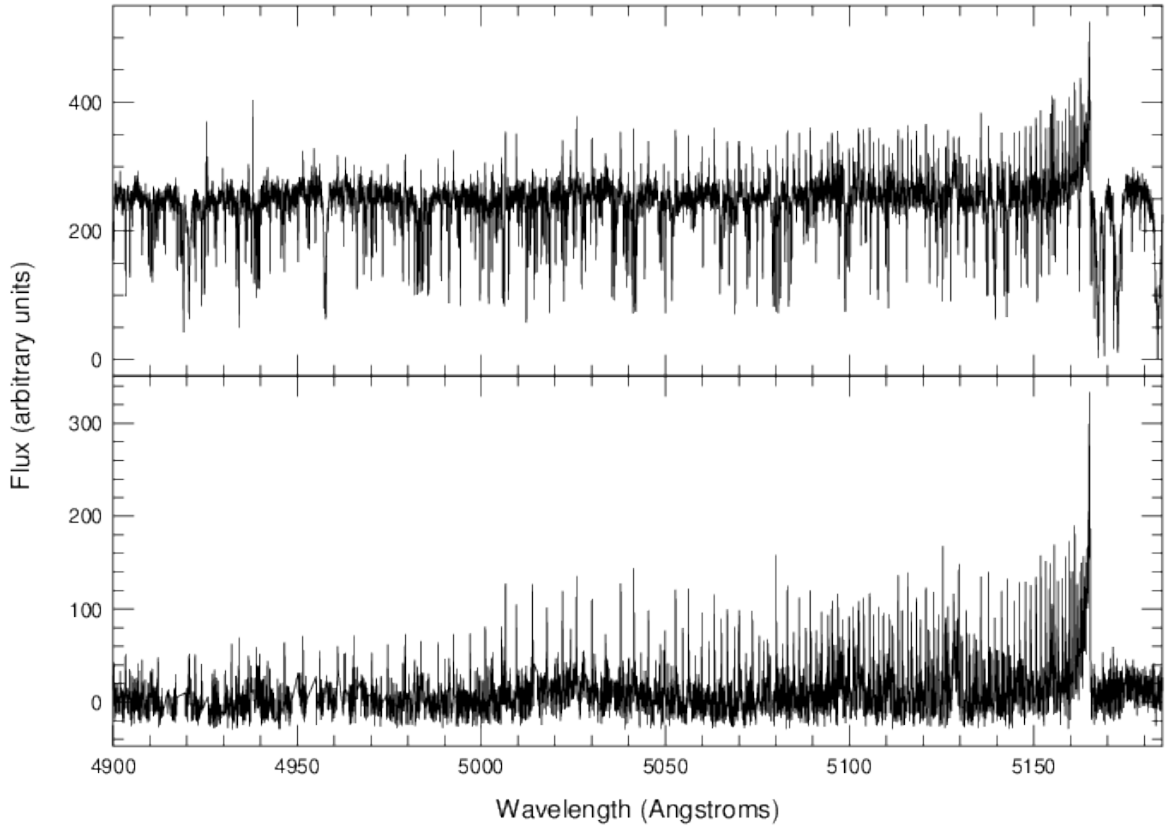}
	\caption{The top panel shows the combined orders for the C$_2$ $\Delta v = 0$ band from 2018 September 15 prior to removal of the dust. It is apparent in this spectrum that there is indeed C$_2$ in GZ's spectrum. However, quantifying it is not possible until the solar spectrum reflected from the dust is removed. The pure emission spectrum is shown in the bottom panel.  While the resultant pure emission spectrum is not super high signal/noise, it is sufficient for our purposes. \label{c2spectra}}
\end{figure}

The next most dominant features in the optical spectra of comets, after CN, are the C$_2$ d\,$^3\Pi_g$ -- a\,$^3\Pi_u$ (Swan) bands.  These are the molecular emission bands that are used to define the class of depleted comets.  The $\Delta v = 0$ Swan bands cover more than 5 orders of the Tull spectrograph.  In order to combine the orders, taking into account the relative sensitivities of the different orders, we used the procedure outlined in \citet{neco19}.  In addition, we needed to remove the solar spectrum that is reflected by the dust before we could determine how much C$_2$ was present and the distribution of the lines. This procedure is also described in Nelson and Cochran.  Figure~\ref{c2spectra} shows the combined $\Delta v = 0$ Swan bands for GZ before and after dust removal. The reflected dust spectrum is quite strong relative to the C$_2$ emission spectrum; the resultant dust-removed, pure emission spectrum has a moderate signal/noise.

\begin{figure}[h!]
	\centering
	\includegraphics[scale=0.8,keepaspectratio]{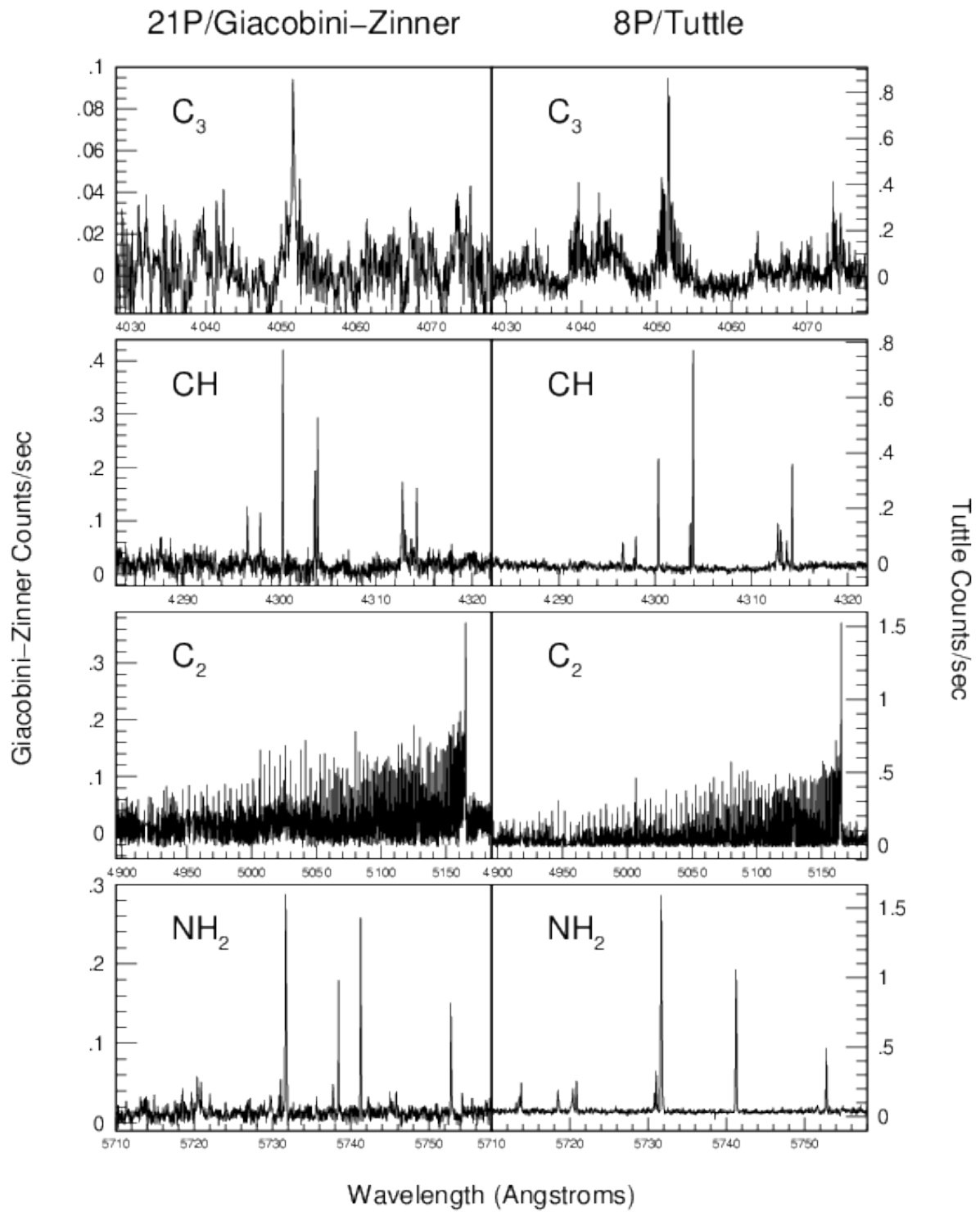}
	\caption{Spectra of four molecules for GZ (left column) and Tuttle (right column) are shown with each molecule scaled to show the full range of the counts in each panel. The molecule is labeled in each box.  The signal/noise is generally much higher for Tuttle than for GZ.
	 The Y-axis values are the count rates for GZ on the left side,  while the Tuttle values are on the right side. \label{allmols}}
\end{figure}

In addition to CN and C$_2$, we detected molecular emissions due to the C$_3$ (A\,$^1\Pi_u$ -- X\,$^1\Sigma_g^+$), CH (B\,$^2\Sigma^-$ -- X\,$^2\Pi$), and NH$_2$ (\~{A}\,$^2\mathrm{A}_1$ -- \~{X}\,$^2\mathrm{B}_1$) bands.  Figure~\ref{allmols} shows spectra of these bands plus C$_2$ for both GZ and Tuttle.  For C$_2$, this is the combined spectrum over 5 orders, as described above.  For the other species, this figure shows a single order, with the solar (dust) spectrum removed as needed.  The GZ and Tuttle spectra are scaled so that the strong features peak at the top of each panel in order to allow a direct comparison of the bands. Clearly, all of these features show up in both comets.  Note that one cannot simply compare count rate of one
species to that of another species because we have not calibrated out
the relative response of the orders.  The redder orders are closer to the blaze 
wavelength of the grating than the blue orders. Also, the quantum efficiency
is higher at the redder end of the spectrum. However, as we explain
below, since we were set up precisely the same for both comets, one
can compare the count rates of a single species between the two comets.

For C$_3$, the strongest peak at $\sim4050$\,\AA\ is readily apparent in both comets.  Less obvious in GZ are the sub-bands at $\sim$4045 and $\sim$4075\AA.   Both are detected in GZ, but are difficult to represent in the plot because of the noise. Examined on the computer, they are 5--10$\,\sigma$ detections.

\begin{figure}
	\centering
	\includegraphics[width=\linewidth,keepaspectratio]{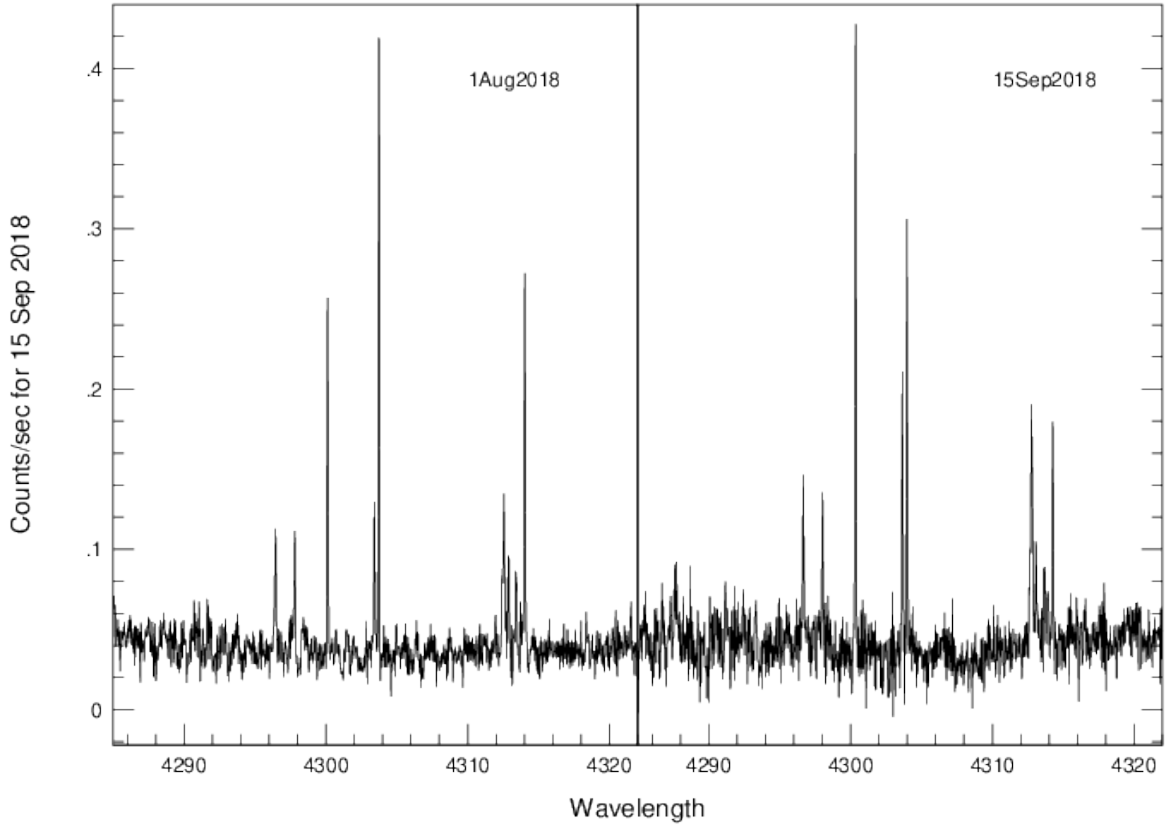}
	\caption{The CH spectra of GZ from 2018 August 1 (heliocentric radial velocity = -11.64 km sec$^{-1}$) and 2018 September 15 (heliocentric radial velocity = +1.65 km sec$^{-1}$ ) are shown.  Note the changes in the relative line strengths from the Swings effect.  The distribution of the line strengths from 2018 August 1 is a much better match to the Tuttle CH spectrum (heliocentric radial velocity = -10.78 km sec$^{-1}$) shown in Figure~\ref{allmols}, as would be expected. \label{ch} }
\end{figure}

\begin{figure}
	\centering
	\includegraphics[scale=0.7,keepaspectratio]{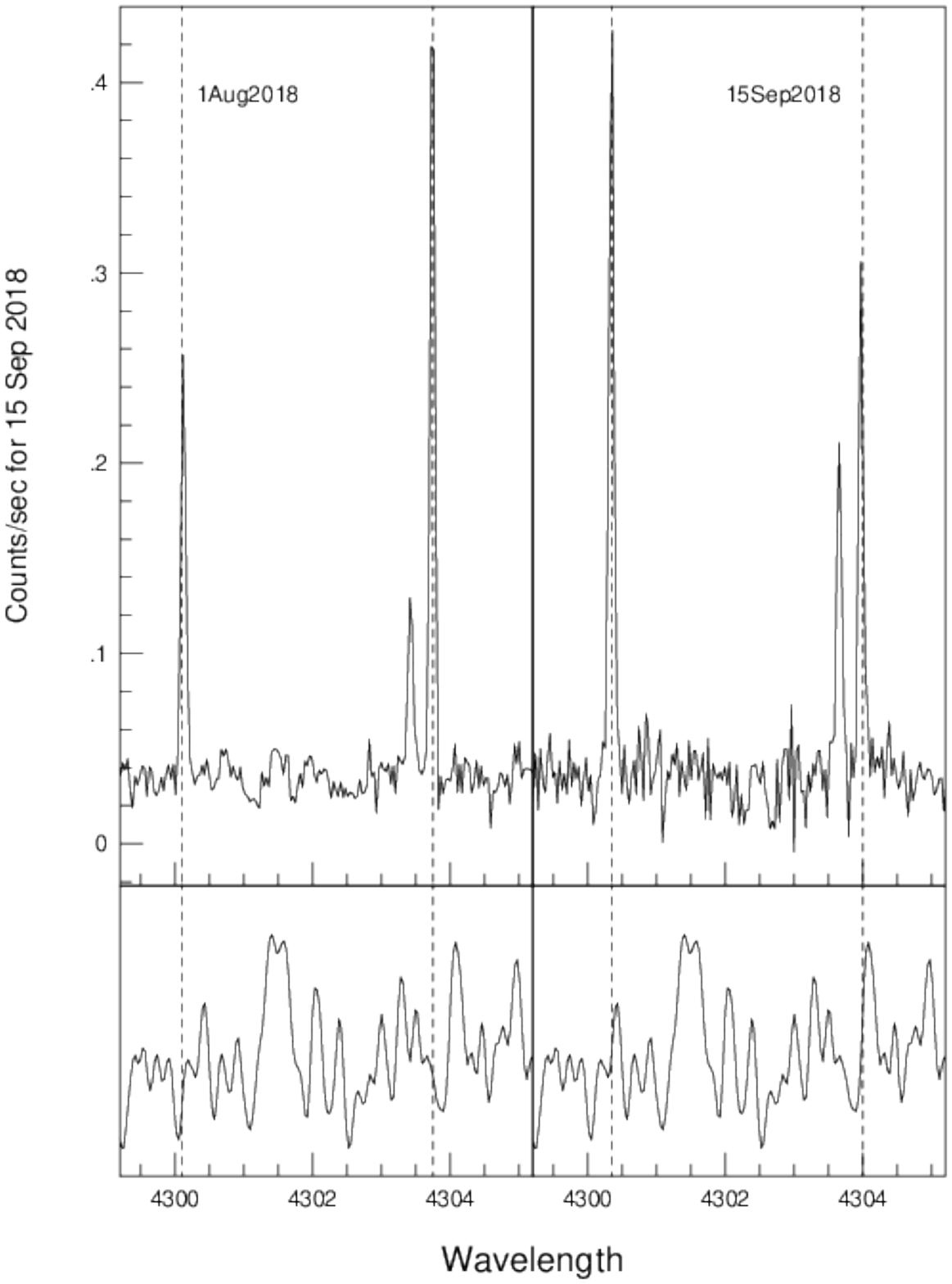}
	\caption{This figure shows a blow-up of the spectra in Figure~\ref{ch} in the region around the 4300 and 4304\AA\ lines. Underneath each panel is the solar spectrum obtained with the same spectrograph (the same solar spectrum is under each comet spectrum). All three spectra are on the telescope restframe. Dotted lines run from the two strongest lines in the comet spectra down through the solar spectra. Note the ratio of the solar flux at the locations of the two strong cometary lines. This explains why the GZ data at a very different heliocentric velocity have different line strength ratios than the Tuttle data.
	 \label{chswings} }
\end{figure}

CH is detected clearly in both comets, though the line ratios are very different.  The ratio of the 4300\AA\  line to the 4304\AA\ line changes significantly, with the bluer line being stronger for GZ and weaker for Tuttle. CH suffers the most Swings effect \citep{swingseffect} of all our observed species, because there are very few lines and they fall at the same wavelengths as the solar G-band, which is mostly caused by CH in the Sun's atmosphere.  Since the cometary emissions are due to resonance fluorescence, the Swings effect describes the variation in emission strength from comet species which are Doppler shifted into and out of solar absorption features. Indeed, inspecting spectra in the region of CH for all of our nights of observations shows that the night of 2018 September 15, the night in figure~\ref{allmols}, is the only night that has a significantly different ratio of 4300\AA\ to 4304\AA\ line strengths than the Tuttle observations.  For 2018 Jul 30 through 2018 Aug 20, the heliocentric radial velocity for GZ ranged from -12.17 to -7.00 km sec$^{-1}$, while Tuttle was observed with a heliocentric radial velocity of -10.78 km sec$^{-1}$.  In contrast, GZ on 2018 Sep 15 had a heliocentric radial velocity of +1.65 km sec$^{-1}$. Not only is the sign the opposite of the other nights' observations, but it is very close to 0 km sec$^{-1}$, where the CH in the Sun's atmosphere would have the greatest Swings effect. Figure~\ref{ch} shows the CH spectra for GZ on both 2018 Aug 1 and 2018 Sep 15.  Figure~\ref{chswings} shows the region around the 4300 and 4304\AA\ lines blown up with the solar spectrum underneath.  All spectra are at the telescope restframe. Note the ratios of the solar flux at the wavelengths of the two strong CH lines. The relative line strengths of the 2018 Aug 1 data are a much closer match to the relative line strengths of Tuttle shown in Figure~\ref{allmols}.  CH is the only molecule to show a strong Swings effect because the band has so few lines and sits on the G-band.  The Swings effect has a slight effect on the appearance of the CN band shown in Figure~\ref{CN} and extremely little effect on C$_3$, C$_2$ or NH$_2$. 

NH$_2$ is also clearly detected in both comets. GZ seems to show some extra lines at around 5738\AA.  We do not have an explanation for these extra lines. These lines could possibly be from another, unidentified species.

\begin{figure}
	\centering
	\includegraphics[width=\linewidth,keepaspectratio]{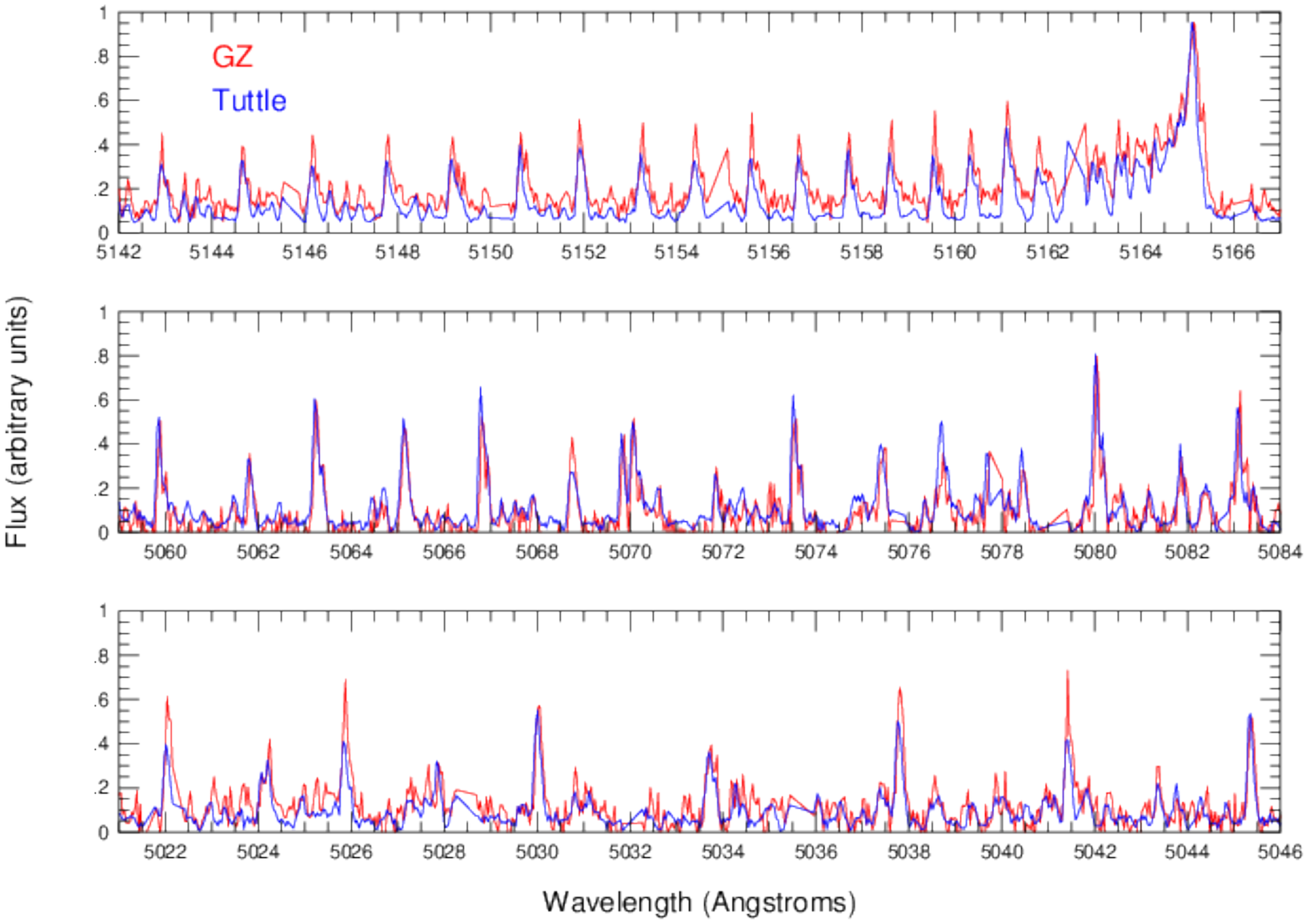}
	\caption{A comparison of three different regions of the C$_2\, \Delta v=0$ band is shown. In the upper panel, the GZ and Tuttle spectra are normalized at the bandhead.  Note that the GZ spectrum in this region is much brighter relative to the bandhead than is the Tuttle spectrum. In the middle panel, we show the region from 5059--5084\AA, normalized at 5080\AA. The line ratios for GZ and Tuttle are very similar, with a few of the Tuttle lines being slightly stronger. The bottom panel shows the region from 5021--5046\AA, normalized at 5045\AA.  In this panel, many of the stronger (P-branch) lines for GZ are relatively much brighter than for Tuttle.}
	 \label{comparec2} 
\end{figure}

C$_2$ is detected in both comets and shows, perhaps, the most interesting inter-comet trend.  C$_2$ is a homonuclear molecule.  As a result, it is not easily de-excited, causing very high J-level lines to be populated.  Indeed, \citet{coco02atlas} showed that lines with J-levels as high as 109 were detected in spectra of 122P/de Vico. Higher J-level lines can not be observed because they are at the same wavelengths as the bandhead of the next bluer band sequence (the $\Delta v=1$ sequence).    Figure~\ref{comparec2} shows three segments of the C$_2$ spectra of Tuttle and GZ overplotted, with each segment of spectrum normalized at the wavelength noted in the caption. The bandhead (top panel) is much stronger relative to the nearby lines for Tuttle than for GZ. In the middle panel, the relative strength of the lines is comparable between the two comets, with Tuttle sometimes stronger than GZ relative to the surrounding lines. In the bottom panel, the stronger lines belong to GZ.  

\begin{figure}
	\centering
	\includegraphics[width=\linewidth,keepaspectratio]{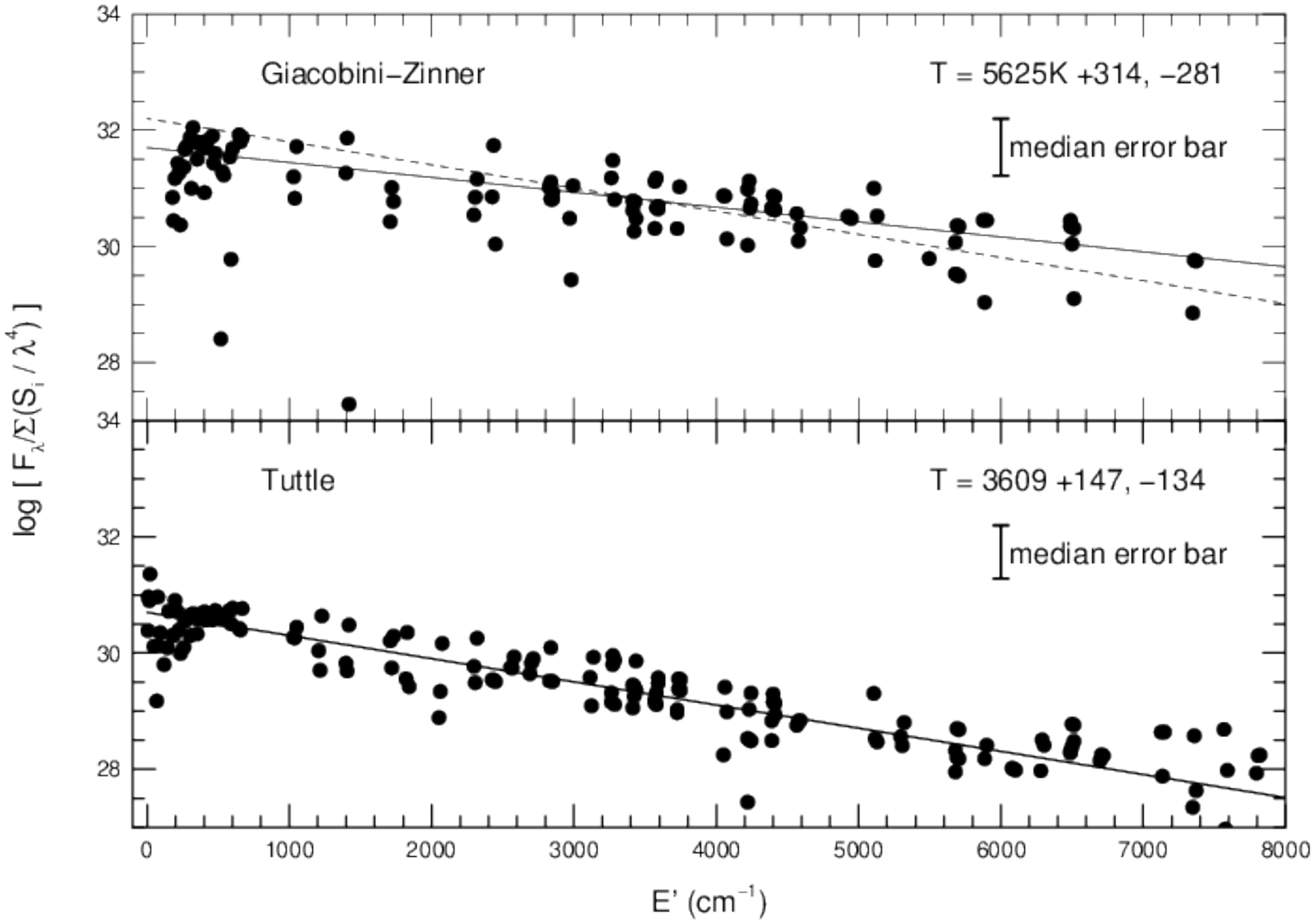}
	\caption{The Boltzmann diagrams for both comets are shown. The temperature is derived from 1/slope of the linear fit shown.  An error bar that is the median of the individual data point errors is shown in each panel.  Some of the points have error bars not much larger than the data point, while others fill much of the panel. Larger error bars generally are for the weakest lines. In the upper panel, there is a dotted line shown that is the slope of the Tuttle fit translated upwards to the GZ data.  One can argue that is too would be a reasonable fit for the GZ data.}
	 \label{Boltzmann} 
\end{figure}

As shown in \citet{neco19}, these differences are due to different rotational (not physical) temperatures in these two comets.  Using the Boltzmann diagram approach described in Nelson and Cochran,  the C$_2$ gas in Tuttle has a rotational temperature considerably cooler than GZ ($5625 +314, -281$K for GZ and $3609 +146, -134$K for Tuttle). These are 1$\sigma$ errors and represent propagated errors from the spectral order blaze removal all the way through the final fit. The Boltzmann plots are shown in figure~\ref{Boltzmann}. Inspection of Figure~\ref{Boltzmann} shows that the scatter in the data can allow for reasonably large changes in the slopes, and hence the temperatures. Indeed, we show the Tuttle slope on the GZ data and it is a convincing fit except at the lowest J-levels, which mostly drive the higher temperature for GZ. 
The difference in temperature is probably too large to be explained by the difference in heliocentric distances of the two comets (1.02 AU for GZ and 1.14 AU for Tuttle at the time of their respective observations).  Most of the difference can be explained as being due to the difficulty in accurately removing the continuum, which is a significant fraction of the signal for GZ (see Figure ~\ref{c2spectra}; note the width of the continuum) and is a lower fraction for Tuttle, coupled with the weakness of the lines in GZ making measurement of the slope in the Boltzmann plot  uncertain.  The temperature  derived from a Boltzmann plot is proportional to 1 over the slope. The slope is of order 10$^{-4}$, so small errors in the slope result in larger errors in the temperature.  To test the effect of improper removal of the continuum, we altered the derived scaling factor of the color-matched solar spectrum by $\pm10\%$ and recalculated the Boltzmann temperature. We find that a 10\% weighting error changes the temperature by 200K, in addition to the propagated errors.  Such an error of the weighting affects the weaker lines more than the stronger lines because the continuum error is a larger percentage of the strength of the integrated flux, thus causing more scatter in the Boltzmann plot. While this does not account for all of the difference in the temperatures of the two comets, it does explain some of the difference. However, as pointed out above, and shown in Figure~\ref{comparec2}, we do see real differences in the relative strengths of lines between the comets, so some temperature difference must be real. Most significant in these plots is that both of the spectra from these comets can be modeled with a single rotational temperature, in contrast with the dual temperature optocenter spectra of 122P/de~Vico and 153P/Ikeya-Zhang modeled in Nelson and Cochran.    Since the lower J-levels drive the temperature for GZ, we tried fitting GZ with two temperatures. This did not improve the fit, so we adopt the simplest conclusion that there is only 1 temperature. An alternative explanation for the temperature difference between GZ and Tuttle could be that it is the result of a different C$_2$ parent for GZ than for Tuttle.

Thus, from Figure~\ref{allmols}, we have been able to show that a depleted comet  still has some (sometimes small) amount of all of the normal species observed in typical comets. Also, the relative line strengths can generally be explained by the Swings effect or different rotational temperatures.  The depletion appears to be a true lower amount of some species relative to CN,
 rather than some atypical distribution of lines. In order to quantify this, we measured the count rates of  parts of each of the various bands for both comets and derived the ratio for each molecule.  Unlike the photometry results of \citet{ahetal95} or the low resolution spectra of \citet{coetal2012}, we did not necessarily include the whole band because often there were long stretches of continuum between lines that add noise
without additional signal.  For C$_3$, we could only measure the central strong part of the band.  Error bars were derived by resetting the continuum several ways and remeasuring the band. Table~\ref{ratios} gives the  count rate ratios of these species between the two comets. We compare with ratios of production rates of \citet{ahetal95}, normalizing their values by setting the CN ratio of these two comets equal to 1.  Values from \citet{coetal2012} are comparable.  
Our C$_2$ ratio is quite a bit higher than that derived by A'Hearn et al., but our C$_3$ ratio is very similar to theirs.  The high C$_2$ value is likely due to the fact that we included only the low J-values of the (0,0) band for our count rate measurements, in the region with the strength differences described above, and not the whole $\Delta v=0$ complex. 
Inspection of Figures~\ref{allmols} (C$_2$ panel) and \ref{comparec2} shows that the lines in this region
are stronger in GZ relative to the bandhead than for Tuttle. This thus causes
our C$_2$ measurement to be high.
We also found that CH and NH$_2$ were
depleted by comparable amounts, though perhaps they are
not quite as depleted as C$_3$.  This is the first such 
measurements of CH depletion in GZ of which we are aware.

\begin{table}[h!]
	\centering
		\caption{Ratio of count rates of species for GZ/Tuttle \label{ratios}}
	\begin{tabular}{cccc}
	\hline
	Species & Ratio & Bandpass & A'Hearn et al. \\
	        &       & measured (\AA) & ratio \\
	\hline
	CN & 1.02$\pm$0.01 & 3860.0-3882.0 & 1 \\
	C$_3$ & 0.11$\pm$0.01 & 4049.0-4056.0 & 0.15 \\
	CH$^*$ & 0.27$\pm$0.02 & 4296.0-4315.5 & \\
	C$_2$ & 0.43$\pm$0.03 & 5131.0-5166.0 & 0.16 \\
	NH$_2$ & 0.28$\pm$0.1 & 5727.5-5753.5 &\\
	\hline
\multicolumn{4}{l}{* Using the CH observations of 30 Jul, when } \\
\multicolumn{4}{l}{\hspace{1em}the heliocentric radial velocity of GZ was} \\
\multicolumn{4}{l}{\hspace{1em}similar to Tuttle's (important for the Swing's} \\
\multicolumn{4}{l}{\hspace{1em}effect), we derive a CH ratio of 0.30$\pm$0.15,} \\
\multicolumn{4}{l}{\hspace{1em}essentially the same as on 15 Sep, with its} \\
\multicolumn{4}{l}{\hspace{1em}different heliocentric radial velocity.}
	\end{tabular}
\end{table}

We did not try to calibrate these ratios into the more common units of production rates.  As long as we are comparing the same features in two comets at comparable heliocentric and geocentric distances, the fluorescence efficiencies should be comparable, as should the outflow velocities. In addition, the observations were obtained over many hours in most cases, so applying a flux calibration  and extinction correction is difficult.  We used the same exact setup for both the GZ and Tuttle observations, so that all features fall on exactly the same part of the detector. There was no cloud cover during the 2018 Sep 15 GZ observations, nor for the Tuttle observations. Both sets of observations were obtained over comparable airmass ranges ($<1.6$ airmasses for the spectra for both comets). Thus, the relative response of the detector at different spectral orders should be the same. The only potential differences would be differences in seeing, though with an extended source this is not a big effect on the flux. Without an absolute calibration, we expect that the ratios of the GZ count rates to the Tuttle count rates  should be accurate to within about 25\%, though formal errors from multiple measurements actually show lower uncertainties.  However, inspection of the values in Table~\ref{ratios} show that we see much larger variations from species to species than can be accounted for by our lack of calibration.  We conclude that CH and NH$_2$ have comparable depletions, C$_2$ is slightly less depleted (but see discussion above) and C$_3$ is the most depleted.

\section{Discussion}
 We have shown that C$_2$/C$_3$ depleted comets look remarkably similar to the ``typical" comets for species such as CN, that generally do not show the depletions. The spectra also do not show significantly different {\it relative line ratios} in the depleted species (such as C$_2$) for GZ when compared with the line ratios of the same species in Tuttle.   As shown above, \citet{ahetal95} showed that CN/H$_2$O was very similar for GZ and Tuttle. The depleted species we detected are depleted by much more relative to CN than the differences in this ratio. Therefore, this implies that the depleted species we detected were also depleted relative to water.

If the solar nebula was more or less homogeneous when the comets formed, and the cometary compositions therefore were also homogeneous, then much lower abundances of some species might come about because of loss of the parent volatile ice by sublimation from the nucleus.  One would then expect that comets that have been near the Sun over longer periods of time and/or more often would be more depleted. Thus, we would expect Jupiter Family comets to be depleted, but long period comets to be unlikely to be depleted. While a higher percentage of Jupiter Family comets (37\%) are depleted than long period comets (19.5\%) \citep{coetal2012}, we see comets that have been in the inner Solar System frequently (e.g. 2P/Encke) that are typical, and comets that have rarely been near the Sun being depleted.

If not the result of dynamical evolution, then the depletions must be from differences in formation.  Long period comets generally enter the inner Solar System from the Oort cloud, having first formed in the giant planet region and then being perturbed outward to the Oort Cloud \citep{donesetal2004}. Jupiter Family comets formed in the Kuiper Belt and are perturbed into the inner Solar System from the scattered disk \citep{duncanetal2004}. This scenario would explain a difference of depletion type if all of the depleted comets were from one reservoir.  However, that is not the explanation these dynamical scenarios would suggest unless the two reservoirs have been mixed at some time and the original formation regions are thus intertwined.  There is a growing consensus that such mixing is likely fairly common (\citet{donesetal2015} and references therein).

The existence of depleted comets in both reservoirs implies that the formation regions were not totally homogeneous.  Pockets with lower quantities of the parents of C$_2$ and C$_3$ must have existed to form the depleted comets.  Dynamical studies to determine how materials were mixed are beyond the scope of this paper.

\citet{rothetal20} obtained complementary IR observations of GZ during a similar period as our observations.  That study targeted the hyper-volatiles CO, CH$_4$ and C$_2$H$_6$ along with observations of H$_2$O.  They found that the CO mixing ratio looked like other Jupiter family comets, but that CH$_4$ might be enriched (though they had some caveats to that statement), while C$_2$H$_6$ was depleted.  They found that some of the species were variable, which leads to the question of whether the depleted comets are always depleted or if the depletion is just a time-variable property.  They concluded that the variability is much smaller than the amount of depletion detected.  They also pointed out that GZ has been observed over more than 1 apparition and always shows depletion.  This is consistent with what has been found in the optical.  Therefore, the depletion is a real effect of the whole body and not just observing different regions of the nucleus.  

 \citet{faggietal2019} also observed GZ in the IR around the time that \citet{rothetal20} did.  They detected H$_2$O, C$_2$H$_6$, CH$_3$OH, HCN and CO and derived only upper limits for C$_2$H$_2$, H$_2$CO, CH$_4$ and NH$_3$. Acetylene is presumed to be the parent for C$_2$ and they found that their upper limits of acetylene to water were consistent with the depletion seen in the optical.  \citet{faggietal2019} also saw some variability of the measured species, with ethane and methanol being depleted sometimes and enhanced at others. However, as with \citet{rothetal20}, the variability is much smaller than the depletion seen in
C$_2$ and other depleted species.

Interestingly,  observations of Tuttle in the IR showed it to also be hydrocarbon poor, though it looks quite typical in the optical \citep{bonevetal2008,koetal8P}.  This suggests that carbon chain depletion may not be tracing hydrocarbons.

In summary, depleted comets do not totally lack the normal species that exist in larger amounts in typical comets.  The small amounts of some species relative to typical comets behave normally when the comet is heated and sublime. There is just not as much of these species' parents to sublime.

\acknowledgements
The observations reported in this paper were obtained at The McDonald Observatory, operated by The University of Texas at Austin.  This work was supported by NASA Grant NNX17A186G.

\end{document}